\documentclass[twocolumn]{aastex63}

\def\eang{$\; \buildrel ^\circ \over . \;$}
\def\ang{\kern-0.3em \eang\kern-0.3em}

\usepackage{amsmath}
\received{May 25, 2020}
\revised{July 9, 2020}
\accepted{XXX}%\today}
\submitjournal{ApJ}

\shorttitle{Off-axis refreshed shocks}
\shortauthors{Lamb, Levan, \& Tanvir}
\begin{document}

\title{GRB\,170817A as a Refreshed Shock Afterglow viewed off-axis}

\correspondingauthor{Gavin P. Lamb}
\email{gpl6@le.ac.uk}

\author[0000-0001-5169-4143]{Gavin P. Lamb}
\affiliation{School of Physics and Astronomy, University of Leicester, University Road, LE1 7RH, UK}

\author[0000-0001-7821-9369]{Andrew J. Levan}
\affiliation{Department of Astrophysics, Radboud University, 6525 AJ Nijmegen, The Netherlands}

%\collaboration{1}{(AAS Journals Data Scientists collaboration)}

\author[0000-0003-3274-6336]{Nial R. Tanvir}
\affiliation{School of Physics and Astronomy, University of Leicester, University Road, LE1 7RH, UK}
%\affiliation{AAS Journals Associate Editor-in-Chief}
%\nocollaboration{1}

%\altaffiliation{AASTeX v6+ programmer}

%\collaboration{1}{(LaTeX collaboration)}



\begin{abstract}
Energy injection into the external shock system that generates the afterglow to a gamma-ray burst (GRB) can result in a re-brightening of the emission.
Here we investigate the off-axis view of a re-brightened refreshed shock afterglow.
We find that the afterglow light-curve, when viewed from outside of the jet opening angle, could be characterised by a slow rise, or long-plateau, with a maximum flux determined by the total system energy.
Using the broadband afterglow data for GRB\,170817A, associated with the gravitational wave detected binary neutron star merger GW170817, we show that a refreshed shock model with a simple top-hat jet can reproduce the observed afterglow features.
We consider two particular refreshed shock models: a single episode of energy injection; and a period of continuous energy injection.
The best fit model parameters give a jet opening angle, for our first or second model, respectively, of  $\theta_j=5$\ang$2^{+1.1}_{-0.6}~$\,or\,$~6$\ang$3^{+1.7}_{-1.1}$, an inclination to the line of sight $\iota=16$\ang$0^{+3.4}_{-1.1}~$\,or\,$~17$\ang$8^{+4.5}_{-2.9}$, an initial on-axis isotropic equivalent kinetic energy $E_1 = (0.3^{+3.5}_{-0.3}~$\,or\,$~0.5^{+6.7}_{-0.2})\times10^{52}$\,erg and a total/final,  on-axis isotropic equivalent refreshed shock energy $E_{\rm total}=(0.42^{+5.6}_{-0.4}~$\,or\,$~1.26^{+18.2}_{-0.7})\times10^{53}$\,erg.
The first model fitting prefers an initial bulk Lorentz factor $\Gamma_{0,1}<60$, with a comparatively low central value of $\Gamma_{0,1}=19.5$, indicating that, in this case, the on-axis jet could have been a `failed-GRB'.
Alternatively, our second model is consistent with a bright GRB for an on-axis observer, with $\Gamma_{0,1}=162.2^{+219.7}_{-122.1}$.
Due to the low-Lorentz factor and/or the jet opening angles at $\theta_j\sim\iota/3$, both models are unable to reproduce the $\gamma$-ray emission observed in GRB\,170817A, which would therefore require an alternative explanation such as cocoon shock-breakout.

\end{abstract}



\section{Introduction}

Gamma-ray bursts (GRBs) are the result of energy dissipation \citep[see][for a review of the prompt GRB emission]{peer2015} within the ultra-relativistic jets launched during the core-collapse of massive stars or the merger of neutron stars \citep[see][for a review of GRB central engines]{fryer2019}.
Due to the relativistic beaming of the prompt emission, cosmological GRBs are typically observed within the jet opening angle, or the beaming angle of the emission.
Observations of the afterglows of GRBs have revealed that in some cases, the external shock responsible for the afterglow emission exhibits a variability that is consistent with the outflow being refreshed \citep[e.g.][]{laskar2015}.

Variability in the afterglows to GRBs provides information about the energetics of a long-acting engine or the jet anisotropy \citep{ioka2005}.
A GRB afterglow's emission is the synchrotron radiation from a decelerating relativistic shock \citep[e.g.][]{sari1998, kobayashi1999} and in the scenario where an engine launches multiple shells with different velocities, the collision of a more energetic but slower shell, catching-up with an initially faster shell that is decelerating, will result in the afterglow re-brightening \citep{rees1998, sari2000}.
Such re-brightening episodes have been observed in a number of GRB afterglows \citep[e.g.][]{dai2001, granot2003, berger2005, fox2005, malesani2007, liu2008, marshall2011, melandri2014, vaneerten2014, laskar2015, lamb2019b}. 

Indeed, the X-ray afterglows of many GRBs show plateaux on timescales of 10$^2$--10$^4$ seconds \citep{nousek2006, zhang2006},  and as long as $10^5$s in extreme cases \citep{grupe2007}, which is readily interpreted as being due to continued energy injection; and for the afterglows of short GRB\,130603B and GRB\,170817A, excess X-ray emission at very late-times supports a long-lived engine \citep{fong2014, piro2019}. 
Such energy injection can be understood as being due to the central engine activity e.g. long-lived accretion onto the nascent black hole from fall-back of material that fails to escape the explosion \citep{rosswog2007}, or through the long-term tapping of the rotational energy of a rapidly spinning magnetar \citep[e.g.][]{gompertz2014,gompertz2017, beniamini2017, gibson2017, gibson2018}.
Alternatively, refreshing of the shock can arise from a short-lived engine simply due to the emission of shells with a wide range of initial Lorentz factors \citep{zhang2006, beniamini2016}.

As the jets responsible for GRBs have half-opening angles of a few degrees,  typically $\lesssim 8$\ang$0$ for both long and short GRBs \citep{goldstein2016, jin2018, wang2018}, and with a tight central value of $\sim2$\ang$5 \pm 1$\ang$0$ for long GRBs \citep{wang2018}%$\lesssim10$\ang0 \citep{goldstein2016} {\bf and typically $\sim2$\ang$5$} for long-duration GRBs \citep{wang2018} and $\lesssim8$\ang0 for short-duration GRBs \citep{jin2018, wang2018}
, then the majority of the progenitor population will result in GRBs that are pointed away from an observer.
Such off-axis GRBs would result in orphan-afterglows \citep[e.g.][]{granot2002} with a handful of candidate events having been found \citep[e.g.][]{cenko2013, cenko2015, marcote2019}.
Where an afterglow is refreshed, the brightest component of the off-axis emission would depend on the total energy of the refreshed shock outflow.
This is particularly important for orphan afterglow searches \citep[e.g.][]{ghirlanda2015, lambtanakakobayashi2018, huang2020}, and for the electromagnetic counterparts to compact binary mergers detected via their gravitational wave (GW) signature  \citep[see][for a review of GW electromagnetic counterparts]{nakar2019}.

The first definitive example of a GRB afterglow observed off-axis is the  X-ray to radio transient first seen at $\gtrsim9$\,days after the
%associated with the coincident 
GW-detected neutron star merger and GRB\,170817A \citep{abbott2017multi, abbott2017}.
As the flux from this transient continued to rise over a period of $\sim150$\,days \citep[e.g.][]{davanzo2018, dobie2018, lyman2018, margutti2018, mooley2018, nynka2018, resmi2018, troja2018}, the origin of the emission was thought to be that from either a choked-jet cocoon, where an unsuccessful jet energises a cocoon of material \citep{murguia-berthier2017}; or a structured jet observed off-axis \citep{lambkobayashi2017, lazzati2017}.
The observations continued to be ambiguous until a rapid decline and imaging via very long baseline interferometry (VLBI) revealed a relativistic and narrow jetted outflow \citep{lambmandelresmi2018, lamb2019a, mooley2018b, troja2018, hotokezaka2019, ghirlanda2019}.

The structured jet scenario to explain the origin of the temporal features in the afterglow to GRB\,170817A has subsequently been widely adopted\footnote{However, see \cite{huang2019} who use jet precession to explain the apparent structure, and \cite{gill2019} who highlight an issue in the assumptions made in calculating an afterglow's emission at early times.}.
The isotropic equivalent kinetic energy in the jet core inferred from structured jet models is typically $\sim$few $\times10^{52}$\,erg \citep[e.g.][]{hajela2019, lamb2019a, ryan2019, troja2019GW170817}; this value is high when compared to the cosmological short GRB population, $\langle E\rangle=1.8\times10^{51}$\,erg \citep{fong2015}, suggesting that GRB\,170817A was a highly energetic short GRB, and would have been observed as such to an on-axis viewer \citep{salafia2019}.

Due to their faintness, short GRB afterglows are not particularly well sampled (when compared to the long GRB afterglow population) \citep[see][for reviews of short GRBs]{nakar2007, berger2014}, however, there are at least two examples of short GRBs with strong evidence for a refreshed shock in the afterglow; 
GRB\,050724 \citep{berger2005, malesani2007}, and GRB\,160821B \citep{lamb2019b, troja2019}.
If the majority of the energy in a short GRB outflow is in a slower component that refreshes the afterglow at a late time, then due to the faintness for an on-axis observer, a refreshed shock episode would be typically difficult to detect.
However, for an off-axis observer the orphan afterglow would be dominated by the higher energy of the refreshed shock.
Here we ask the questions:
{\it What would a refreshed shock afterglow look like to an off-axis observer?
Can a refreshed shock afterglow explain the observed features of the GRB\,170817A afterglow?
%What, if any, are the implications for untriggered orphan afterglow searches?
}

In \S\ref{sec:analytic} we describe the analytic estimates for the peak flux and time for an off-axis afterglow from a refreshed-shock scenario. In \S\ref{sec:numerical} we give the details of our numerical afterglow models used to fit the data from GRB\,170817A.
In \S\ref{sec:results} we describe our results. 
And in \S\ref{sec:disc}\,\&\,\ref{sec:conc} we discuss the results and give our conclusions.

\section{Method}
The observed peak flux and time for a GRB afterglow from a top-hat jet structure, when viewed outside of the jet opening angle, can be estimated analytically \citep[e.g.][]{nakar2002}.
To get a more accurate picture of the behaviour of an afterglow for variously positioned observers, and for any jet structure, we can use numerical methods to better produce the expected light-curves \citep[e.g.][]{granot2002, vaneerten2010, lambkobayashi2017, ryan2019, salafia2019}.

Here we describe our method for estimating the peak flux for the two outflow components considered for an off-axis observed top-hat jet with a single refreshed episode.
We further describe the numerical  models used to explore the parameter space required to fit the observed off-axis afterglow to GRB\,170817A.

\subsection{Analytic peak afterglow estimates}\label{sec:analytic}
We estimate the off-axis peak flux and peak time from a conical outflow with a simple top-hat structure i.e. a uniform energy and velocity within an angle from the central axis until the half-opening angle $\theta_j$, where we assume the energy and velocity drop to zero.
This is the simplest, non-spherical, approximation for the expected outflow structure.

For a refreshed shock afterglow, the light-curve could exhibit a second peak;
the first peak from the onset of deceleration for the initial shell or outflow, assuming the observed frequency $\nu$ is above the characteristic synchrotron frequency $\nu_m$, and a second peak following the collision of the first and second shells, assuming that the second shell has a higher kinetic energy, $E$ than the first \citep{rees1998, sari2000, granot2003}.
To estimate the peak time and flux of these events for an off-axis observer we follow the analytic estimates in \cite{lambkobayashi2017}, here the peak flux\footnote{We assume the slow-cooling regime and the cooling frequency is $\nu_c>\nu$, the observed frequency.}, where $\nu>\nu_m$, for an observer at an inclination $\iota>\theta_j$ is,
\begin{multline}
    F_p \sim (2/3)~C(p)~ f(\iota,\theta_j)~\iota^{2(1-p)}(1+z)^{(1-p)/2}\\\times\nu^{(1-p)/2}~E ~n^{(1+p)/4} \varepsilon_B^{(1+p)/4}~ \varepsilon_e^{p-1}~ D^{-2},
    \label{eq:peakflux}
\end{multline}
where $C(p)$ is a coefficient that depends on $p$, $f(\iota,\theta_j)$ is a function of $\iota$ and $\theta_j$ \citep[see][where the expressions are given explicitly and the $2/3$ is just a numerical factor to account for the difference in the synchrotron flux estimation]{lambkobayashi2017},  $E$ is the isotropic equivalent energy for an on-axis observer, $n$ is the surrounding medium particle number density, $\varepsilon_B$ and $\varepsilon_e$ are the microphysical parameters that define the fraction of energy in the magnetic field and electrons respectively, $p$ the relativistic electron distribution index, and $D$ the luminosity distance to the source.
Equation \ref{eq:peakflux} is valid for $\iota\geq\Gamma^{-1}$.
By considering the sideways expansion of the outflow, the peak flux time can be estimated as \citep[e.g.][]{nakar2002},
\begin{equation}
    t_p \sim 121.3(1+z)~\iota^2 \left[\frac{E}{10^{52}{\rm ~erg}}\right]^{1/3} \left[\frac{n}{0.1 {\rm ~cm^{-3}}}\right]^{-1/3} ~{\rm days}.
    \label{eq:peaktime}
\end{equation}

From equation\,\ref{eq:peakflux}, if we assume that all the parameters are constant at all times with the exception of the total kinetic energy, $E$, that will increase as the shock is refreshed, it is clear that for an off-axis observer the afterglow lightcurve will be dominated by the more energetic refreshed shock emission.
From equation\,\ref{eq:peaktime}, this peak will appear later than that for the emission from the initial outflow.
Where the conditions are right, then the off-axis afterglow lightcurve will exhibit an early rise dominated by the initial outflow followed by a transition to a later, and brighter peak due to the refreshed shock emission.
The delay between these `peaks' is proportional to the difference in the energy as $t_1/t_2 = (E_1/E_2)^{1/3}$, and the flux from either component as $F_1/F_2 = E_1/E_2$, where the subscript indicates the peak time/flux associated with the order of the shells.

\subsection{Numerical afterglow}\label{sec:numerical}
The emission from a forward shock generated during the deceleration of a relativistic outflow is calculated following the method described in \cite{lambmandelresmi2018} with synchrotron self-absorption using the prescription in \cite{lambkobayashi2019}.
A conical outflow with rotational symmetry and defined by the half-opening angle, $\theta_j$, is divided into components following the method in \cite{lambkobayashi2017}.
For each component, eight parameters are required to calculate the forward shock flux due to synchrotron radiation, these are:
$E_1$, the on-axis isotropic equivalent kinetic energy, $\Gamma_0$ the maximum Lorentz factor, $\theta_j$ the half-opening angle, $\varepsilon_B$, $\varepsilon_e$, $n$, $p$, and $\iota$ the inclination of the outflow central axis to the line-of-sight.
By assuming spherical symmetry and via conservation of energy, the parameters, $E_1$, $n$ and $\Gamma_0$ are used to determine the instantaneous Lorentz factor $\Gamma$ and swept-up mass of the decelerating blastwave \citep{peer2012}.
The effects of expansion at the sound speed are included by considering the change in a component's swept-up mass as the outflow solid-angle changes due to the sound speed expansion \citep{lambmandelresmi2018}.

The outflow will penetrate the medium surrounding the central engine and sweep-up matter as it expands resulting in the outflow's deceleration.
A simultaneously launched secondary outflow, or slower shell, that follows the first will travel along the swept-path and catch-up with the initial outflow when $\Gamma_1(r) = \Gamma_{0,2}/2$ here $\Gamma_1(r)$ is the instantaneous Lorentz factor of the leading outflow at a radius $r$ from the central engine, and $\Gamma_{0,2}$ is the maximum Lorentz factor of the slower shell or outflow \citep{kumar2000}; where there is a delay between the outflows, then $\Gamma_1(r)\leq\Gamma_{0,2}/2$.
The energy of the second outflow/shell, $E_2$, is added to the first, $E_1$.
Where $E_2>E_1$, then emission from the shock will result in an observable re-brightening of the afterglow \citep[e.g.][]{rees1998, sari2000, zhang2002, granot2003, lamb2019b}.

To estimate the light-curves from a refreshed shock blastwave, we introduce two new parameters to our method;
$\Gamma_{0,2}$ and $E_2$, the maximum Lorentz factor of the second shell and its kinetic energy.
The dynamics for the decelerating blastwave are calculated via energy conservation.
When the initial blastwave's $\Gamma_1(r) = \Gamma_{0,2}/2$, then the total energy in the shell becomes $E_1 + E_2 = E_{\rm total}$, and the dynamical evolution is calculated assuming this new energy value.
The sudden increase in energy for the decelerating blast-wave results in additional swept-up mass as the shell adiabatically expands to accommodate the increase in energy.
This swept-up mass, along with the change in solid-angle due to sound-speed lateral expansion, is used to determine the instantaneous radius and therefore the timescale.

By calculating the emission from only the forward shock of a blastwave with an increase in energy/mass, we are ignoring the complex conditions within the collision.
As the shells collide, then the collision can be described by either a violent or a mild case.
The nature of the collision depends on the instantaneous Lorentz factor of the two shells at collision and the ratio of the shell energies.
In a violent collision, a strong shock is generated and emission from a reverse shock, that travels back into the catching shell, and a secondary forward shock, that travels into the caught shell should be considered.
These shock conditions and the resulting emission are given in detail by \cite{zhang2002}.

For $E_2/E_1>1$, and $\bar{\Gamma}\gtrsim1.22$, where $\bar{\Gamma}$ is the Lorentz factor of the secondary shell with respect to the co-moving initial shell, and where the spreading radius for the secondary shell is less than the deceleration or collision radius (a so called `thin' shell), then the collision will be violent and emission from the reverse shock should be considered.
However, the reverse shock emission will peak at a very low-frequency and is unlikely to be brighter than the forward shock emission, especially for a thin shell case, unless an additional strong magnetic field is invoked \citep[e.g.][]{harrison2013}; additionally, a reverse shock will peak and rapidly fade before the emission from the associated forward shock for an off-axis observer \citep{lambkobayashi2019}.
The complex emission components for a violent shell collision could therefore result in variability in the lightcurve at about the collision time, and particularly at radio frequencies, however, in all cases of a refreshed shock, the final lightcurve behaviour is determined by the total energy of the system. 
As such, we do not consider here the additional emission components associated with a violent collision.

The simple two shell model that we adopt can be modified so as to ensure that the collision between shells is always mild.
If we assume that a second shell has a velocity distribution where $E(>\Gamma)\propto\Gamma^{1-s}$ \citep[e.g.][]{sari2000}, then the catching shell will inject energy over a finite timescale where the instantaneous energy ratio is always very close to 1 and $\bar{\Gamma}=1.25$ for simultaneously launched shells; 
from Fig. 3 in \cite{zhang2002}, the resultant shock will be mild for a thin secondary shell where $\bar{\Gamma}\lesssim2$ and the afterglow lightcurve can be determined by the forward shock emission alone; such a scenario was used by \cite{laskar2015} for their sample of observed refreshed shock afterglows. %, which showed no evidence for a significant violent collision.
To include this scenario in our models, we require an eleventh free parameter, $s$, the shell mass distribution index.
The dynamics of the decelerating blastwave are then solved with the initial fireball energy and mass until a given collision Lorentz factor, when the energy/mass of the blastwave is increased according to the index $s$ until the total energy of the system, $E_{\rm total}$, is reached and energy injection ceases.

The parameters used to determine the observed light-curve are calculated from the co-moving frame values \citep[see][]{sari1998, wijers1999, lambmandelresmi2018} using the relativistic Doppler factor $\delta = [\Gamma(1-\beta\mu)]^{-1}$, where $\beta$ is the velocity as a fraction of the speed of light and $\mu$ the cosine of the angle from the shock normal to the line-of-sight given by,
\begin{equation}
    \mu = \sin\theta\cos\phi\sin\iota+\cos\theta\cos\iota,
\end{equation}
where $\theta$ is the angle from the central axis to the emission point and $\phi$ is the rotation angle of the point on the emitting surface.

\begin{figure}
    \centering
    \includegraphics[width=\columnwidth]{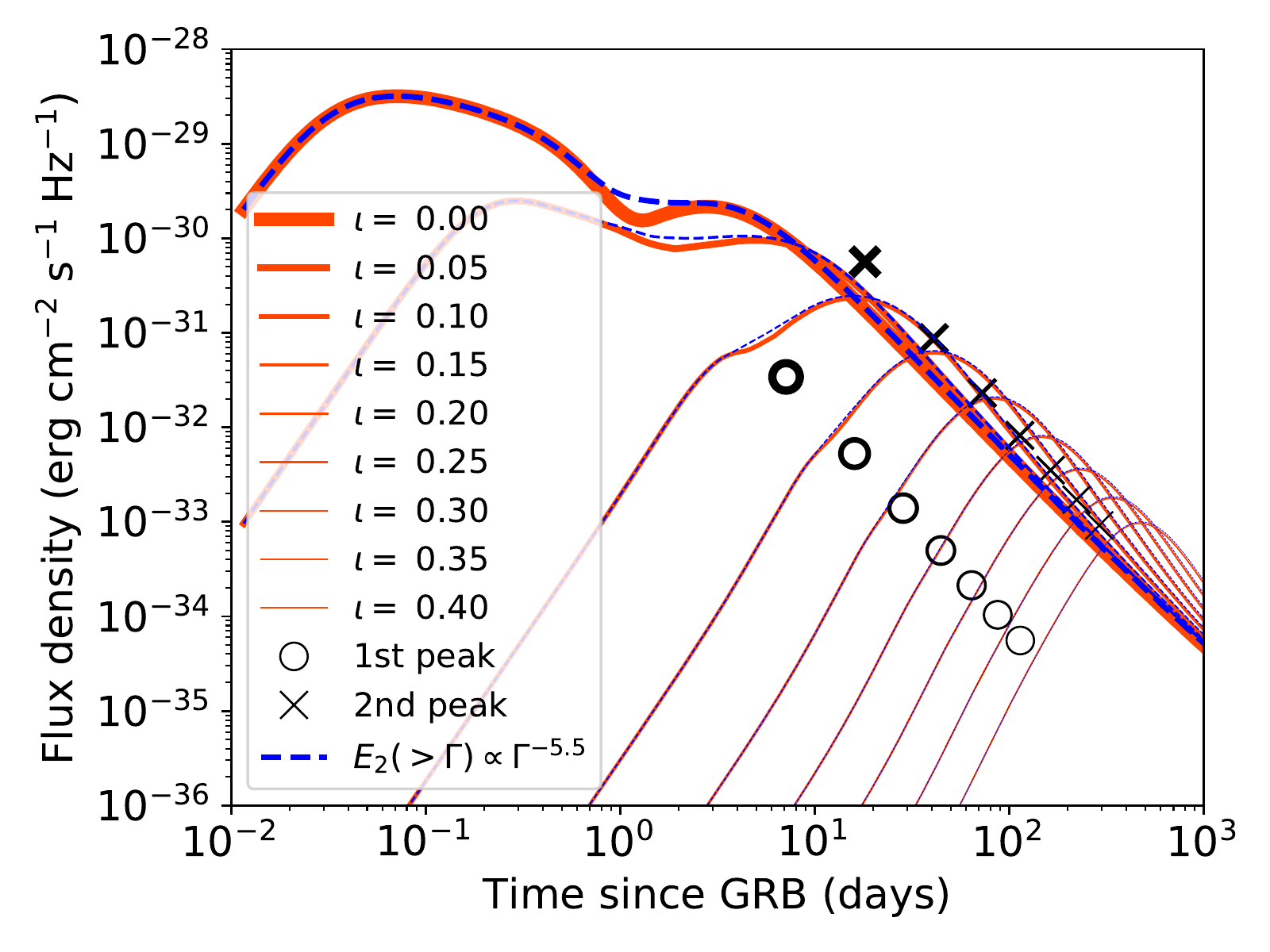}
    \caption{The model forward shock afterglow to GRB\,160821B at R-band using the parameters from \cite{lamb2019b}. The inclination (in rad) of the jet to the line-of-sight is varied from $0.0\leq\iota\leq0.4$ in steps of $\Delta\iota = 0.05$. The refreshed shock component of the afterglow becomes dominant for observers at $\iota\gtrsim2\theta_j$, where $\theta_j = 0.033$\,rad for GRB\,160821B. Black markers show the analytic estimates from equations\,\ref{eq:peakflux}\,\&\,\ref{eq:peaktime} for cases where $\iota>2\theta_j$.
    The blue dashed lines show the case where the refreshing shell has an energy distribution that is proportional to $E_2(>\Gamma)\propto\Gamma^{1-s}$, and the collision Lorentz-factor is chosen so that the refreshed shock has the same peak-time and flux as the discrete shell case.}
    \label{fig:GRB160821B}
\end{figure}

The afterglow to the on-axis observed short GRB\,160821B requires a refreshed shock to successfully accommodate the observed features in the data for this event \citep[see][for a detailed discussion]{lamb2019b}.
To demonstrate how a refreshed shock afterglow would look at various inclinations we use the afterglow parameters for GRB\,160821B and vary the inclination, see Fig.\,\ref{fig:GRB160821B}.
Here the jet has a top-hat, or uniform and sharp-edged structure with parameters:
$E_1 = 1.3\times10^{51}$\,erg, $\Gamma_{0,1}=60$, $E_{\rm total} = 2.1\times10^{52}$\,erg, $\Gamma_{0,2}=24$, $\theta_j=0.033$\,rad, $\varepsilon_B=0.01$, $\varepsilon_e=0.1$, $n=10^{-4}$\,cm$^{-3}$, $p=2.3$, and the inclination $\iota$ is varied from $0.0$ -- $0.4$ radians.
Additionally, the estimates for the peak flux and time for the initial outflow and the refreshed shock from equations\,\ref{eq:peakflux}\,\&\,\ref{eq:peaktime} are shown for the cases where $\iota>2\theta_j$ as black circles and crosses for the initial and the refreshed shock respectively.
To give an indication of how a continuous energy injection episode can affect the lightcurve, we additionally show the case where the refreshing shell has an energy distribution with $s=6.5$ and the collision Lorentz factor is chosen so that the total mass is identical to the discrete energy injection model -- the lightcurve is shown by a blue dashed line.

The temporal phenomenology of these off-axis light-curves can be compared with an observed off-axis afterglow, and vice-versa.
Given the apparent `structure' in the afterglow light-curve for a mildly inclined observer, see the light-curves (red lines) for $\iota\sim0.10$--$0.15$ in Fig.\,\ref{fig:GRB160821B}, we test this model with data from an off-axis observed afterglow.
The only confirmed example of a short GRB afterglow viewed outside of the outflow opening angle is that of GRB\,170817A \citep[e.g.][]{alexander2018, davanzo2018, dobie2018, lyman2018, margutti2018, mooley2018, nynka2018, resmi2018, troja2018, fong2019, hajela2019, lamb2019a}. 
By assuming a simple top-hat jet structure that is refreshed at late-times by a more energetic second component we use a Markov-Chain Monte Carlo (MCMC) to find the best fitting parameters in our model.
To improve efficiency, we vary the resolution of the model depending on the values of key parameters.
The resolution for each outflow is equal to the integer closest to $(2 \theta_j \Gamma_0)^2$, where $\Gamma_0$ is the maximum bulk Lorentz factor of the outflow.
For an off-axis observer, where the system inclination $\iota > \theta_j$, the resolution is reduced proportional to the difference in $\iota-\theta_j$ with a minimum resolution of 100 individual components.
This ensures that the afterglow light-curves are produced without noise from the numerical methods.

We employed \texttt{emcee} \citep{emcee2013} to perform the MCMC with the flat priors given in Table\,\ref{tab:priors}.
The bounds on the prior range for each parameter is set broad to avoid introducing any forced constraints on the fit with the exception of the system inclination and the jet opening angle.
For the inclination, we limit to $0.25\leq\iota\leq0.45$ and flat\footnote{We use a flat $\cos\iota$ distribution to avoid biasing towards lower inclinations.} in $\cos\iota$.
The limits on $\iota$ are those found by \cite{hotokezaka2019} from GW, VLBI and afterglow modelling, and consistent with \cite{mandel2018} using only the GWs, although in mild contention with constraints from the macronova analysis \citep[see][]{dhawan2020}.
For the jet opening angle, despite the tight constraints for a narrow core, or jet half-opening angle $\theta_j$, from the VLBI measurements \citep{ghirlanda2019}, we choose only to limit the opening angle to a range that covers the observed short GRB measured values or limits, $1$\ang$0\leq\theta_j\leq25$\ang$0$ -- noting here the degeneracy in afterglow model fitting with $\iota$--$\theta_j$ \citep{nakar2020}.

\begin{table}
	\centering
	\caption{The \texttt{emcee} prior parameter ranges for the refreshed shock model. The inclination range is limited to that found by \cite{hotokezaka2019}, and the jet opening angle to the observed distribution of short GRB opening angles \citep{fong2015}.
	}
	\label{tab:priors}
	\begin{tabular}{rcl} % four columns, alignment for each
		\hline
		Lower limit & Parameter & Upper limit \\
		\hline
		47 & $\log[E_1$ (erg)] & 54 \\
		$>\log[E_1$] & $\log[E_{\rm total}$ (erg)] & $\log[1000\times E_1$] \\
		2 & $\Gamma_{0,1}$ & 1000 \\
		2 & $\Gamma_{0,2}$ & $<\Gamma_{0,1}$ \\
		0.0175 & $\theta_j$ (rad) & 0.4363 \\
		0.9004 & $\cos[\iota]$ & 0.9689 \\
		-5.0 & $\log[\varepsilon_B]$ & -0.5 \\
		-5.0 & $\log[\varepsilon_e]$ & -0.5 \\
		-6.0 & $\log [n$(cm$^{-3}$)] & 0.0 \\
		2.01 & $p$ & 2.99 \\
		0.00 & $s$ & 100 \\
		\hline
	\end{tabular}
\end{table}

\section{Results}\label{sec:results}

The analytic estimates for the peak flux from the initial and the refreshed shock outflow, as shown in Fig.\,\ref{fig:GRB160821B},
give an adequate approximation for the peak flux and time\footnote{The underestimate in the peak time for the refreshed shock peak is expected, the analytic estimate is based on the time when the Lorentz factor of the decelerating blastwave is equal to $1/\iota$ and does not consider the additional light-travel time due to the changed geometry} compared to the numerical light-curve approximation at all points where $\iota>2\theta_j$, with the exception of the refreshed shock peak flux at $\iota<0.15$ (black crosses).
These exceptions can be understood by considering the beaming cone for the emission from the refreshed shock;
where the shells collide when $\Gamma_1 = \Gamma_{0,2}/2$, which for GRB\,160821B is $\Gamma_1\sim12$ \citep{lamb2019b}.
The refreshed shock gains momentum and coasts before again entering a deceleration phase governed by energy conservation; the afterglow light-curve will flatten or re-brighten before breaking as the outflow starts to decelerate.
The bulk Lorentz factor of the outflow at this time is $\Gamma\sim6$ and where $\Gamma < [\iota-\theta_j]^{-1}$, such as for the two overestimated points at $\iota=0.1$\,\&\,$0.15$, the emission will be beamed towards the observer before the observed peak and so the expression for the peak flux in equation\,\ref{eq:peakflux} is not valid.

\begin{figure}
	\includegraphics[width=\columnwidth]{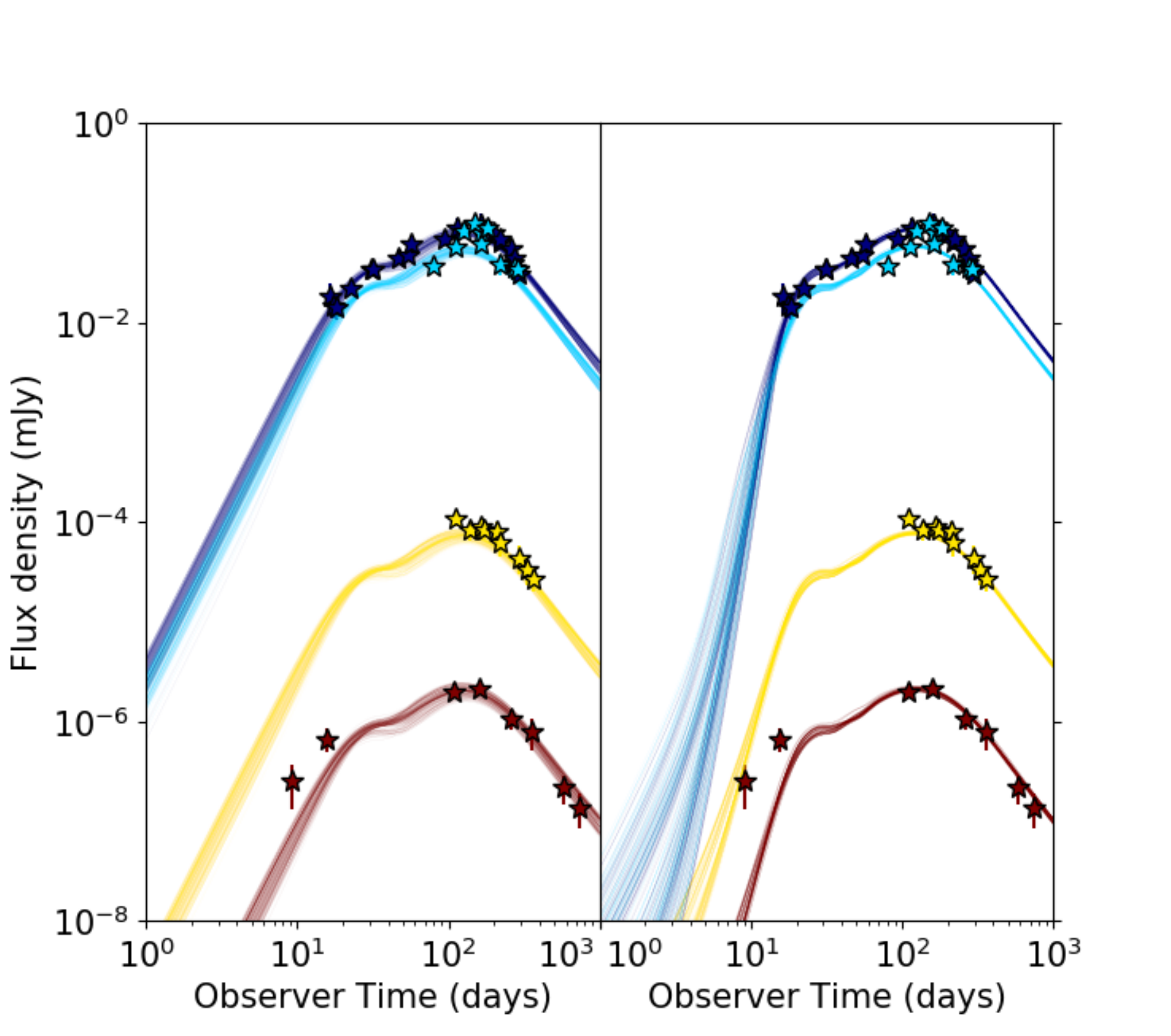}%refreshed-shock-both.pdf}
    \caption{A refreshed shock model fit (Model 1, left -- Model 2, right) to the GRB 170817A radio (3 GHz -- dark blue, 6 Ghz -- light blue), optical (F606W, $\sim5.1\times10^{14}$\,Hz -- yellow), and X-ray (1 keV -- red) data from \cite{alexander2018, mooley2018, fong2019, hajela2019}.
    The coloured lines show the light-curves using 200 randomly sampled posterior distribution parameter sets for GRB\,170817A.} %{\bf The same lightcurves for Model 2 are shown in the inset, where the axis have the same range as the main panel.}}
    \label{fig:GRB170817A}
\end{figure}

To determine if the refreshed shock model can explain the observed features for an off-axis observed afterglow (features that are typically attributed to lateral jet structure) we use an MCMC analysis with the data from the afterglow of GRB\,170817A \citep{alexander2018, mooley2018, fong2019, hajela2019}.
The light-curves from 200 randomly sampled posterior parameters sets are shown in Fig.\,\ref{fig:GRB170817A} with the afterglow data used for the fits.
 The results are shown for two Models:
\begin{enumerate}
    \item A discrete energy injection episode and an assumed mild collision.
    \item A continuous energy injection episode, where the total energy distribution increases as $E(>\Gamma_c)=E_1(\Gamma/\Gamma_c)^{1-s}$, where $\Gamma_c=\Gamma_2/2$ and is the Lorentz factor of the leading shell at collision; during the collision energy is continuously injected until a maximum value, $E_{\rm total}$, is reached.
\end{enumerate}
From these we can see that the refreshed shock model using a simple top-hat jet can recover the observed features, a gradual rising light-curve from the earliest detection at $\sim10$\,days post burst to a peak at $\sim150$\,days, for the afterglow to GRB\,170817A.
The central values and the 16th and 84th percentiles from the posterior distribution for the fit parameters are listed in Table\,\ref{tab:GRB170817A}.

\begin{figure}
	\includegraphics[width=\columnwidth]{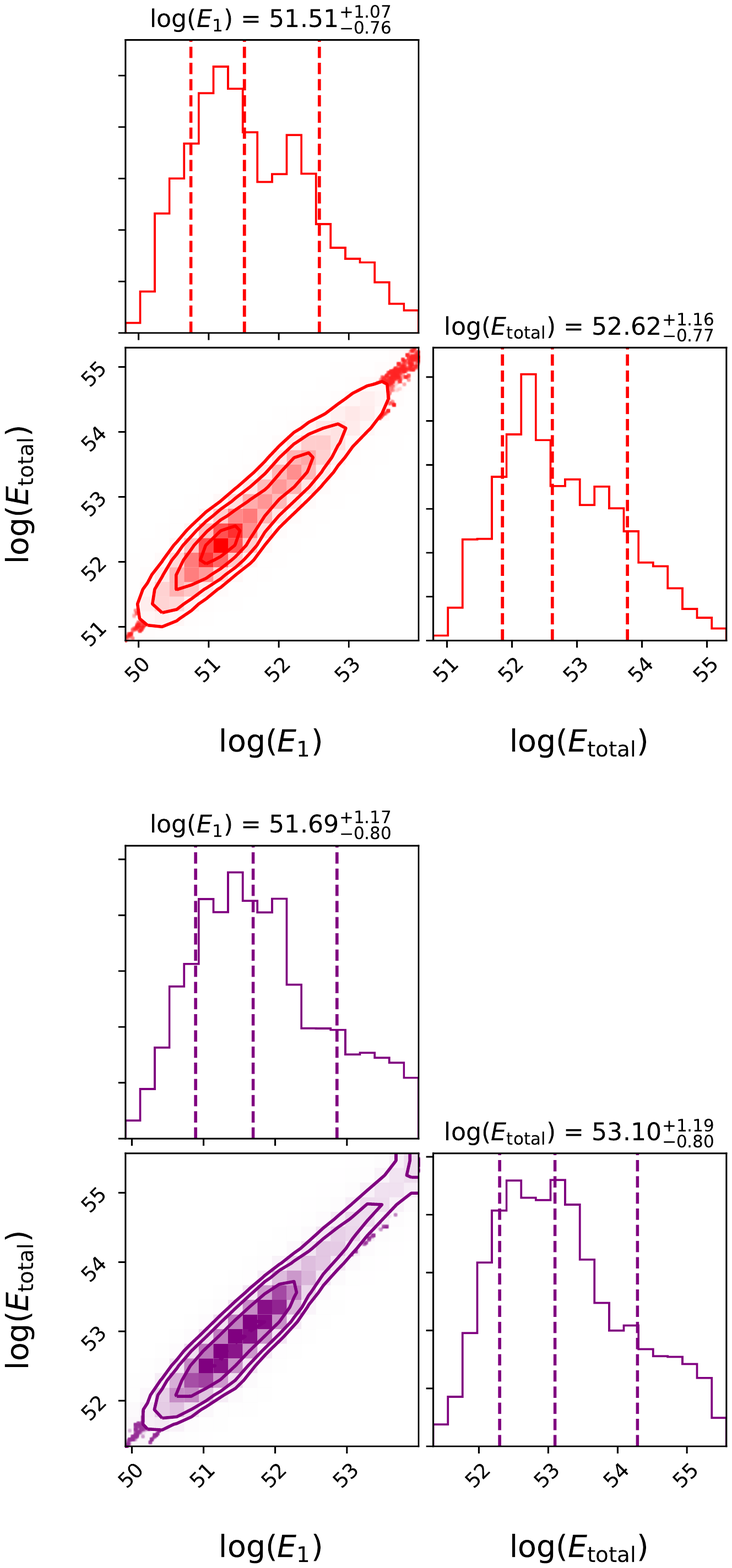}
    \caption{The 2D posterior distribution for the initial outflow energy and the total refreshed shock energy. Vertical dashed lines indicate the 16th, central, and 84th percentiles of the distribution. The parameters demonstrate the expected strong linear correlation, where the the peak flux is proportional to the outflow energy.
    Top panel (red) shows posterior for Model 1.
    Bottom panel (purple) shows posterior for Model 2.}
    \label{fig:energy-corr}
\end{figure}

A tight linear correlation between the kinetic energy, $E_1$ of the initial outflow and the total energy, $E_{\rm total}$ of the refreshed shock is shown in Fig.\,\ref{fig:energy-corr}.
From the analytic estimates, this tight correlation is expected, where the maximum synchrotron flux is $F_{\rm max} \propto E$ \citep{sari1998} and the peak flux then depends on the timescale and the spectral regime \citep{sari1999}.
For an off-axis observer for both the initial and the refreshed shock `peaks', the ratio of the fluxes is equal to the ratio of the energies, e.g. equation\,\ref{eq:peakflux}. 
The ratio of the peak flux to the flux at $\sim15$\,days is $\sim3$, whereas from Table\,\ref{tab:GRB170817A} the ratio of energies is  $\sim12(25)$ for Model 1(2) respectively.
This inequality is similar to the case in Fig.\,\ref{fig:GRB160821B} at $\iota\sim3\theta_j$ and suggests that the observed emission in the afterglow to GRB\,170817A, using the refreshed shock model, is beamed towards the observer before the peak.
The flux is then determined by a post-jet-break observed system where the observer is at $\theta_j<\iota<\Gamma(t)^{-1}$.
For $\iota=0.28(0.31)^{+0.06(0.08)}_{-0.02(0.05)}$, and $\theta_j=0.09(0.11)^{+0.02(0.03)}_{-0.01(0.02)}$\footnote{ Although, we note that where magnetised jets are considered, the opening angle from fits to the GRB\,170817A data result in $\theta_j>0.17$\,rad, with a hollow-core of $\sim0.09$\,rad and a viewing angle $\iota\sim0.37$\,rad \citep{nathanail2020}.} the maximum Lorentz factor for the emission from the light-curve near peak is therefore $\Gamma\leq6.7(8.3)$.
This is consistent with the constraints from the superluminial motion of the centroid in the VLBI observations between 75--230\,days, with an apparent velocity $\beta_{\rm app} = 4.1\pm 0.5$ \citep{mooley2018b}, giving $\Gamma\geq3.6$.

We also see a preference for a narrow jet half-opening angle with $\theta_j = 0.09(0.11)$, or $\sim5(6)^\circ$, for Model 1(2) respectively.
The superluminal motion of the centroid indicates an opening angle $\theta_j\lesssim5^\circ$ \citep{mooley2018b}; 
as $[\iota-\theta_j]>\theta_j$ for our values, this estimate holds for our models and without forcing the narrow jet condition, our results are consistent with the opening angle inferred from the centroid motion.
Where a structured jet model is used to fit the afterglow, the values for the jet core are $\theta_c\sim0.06$, or  $\sim3$\ang$5$ \citep[e.g][]{hotokezaka2019,ghirlanda2019}; however, as the jet core values are the result of light-curve fitting, it is unclear to what extent emission from the region immediately outside of the jet core contributes to the radio centroid images for structured jets \citep[see][who rule out $\gtrsim30^\circ$ outflows to high significance from the centroid images]{ghirlanda2019}.
Both of the VLBI fits utilised core-dominated structured jet models where the inclination and core angle from the fit can depend on the choice of jet structure and how $\theta_c$ is defined \citep{ryan2019}.
Additionally, the inclination $\iota$, of the system from our fits has a preference for lower angles, despite using a flat prior over $\cos\iota$.
The preferred inclination of the system is $\iota=0.28^{+0.06}_{-0.02}$, or $16$\ang$0^{+3.5}_{-1.1}$  for Model 1, and $\iota=0.31^{+0.08}_{-0.05}$, or $17$\ang$8^{+4.5}_{-2.9}$ for Model 2.

\begin{figure*}
	\includegraphics[width=\textwidth]{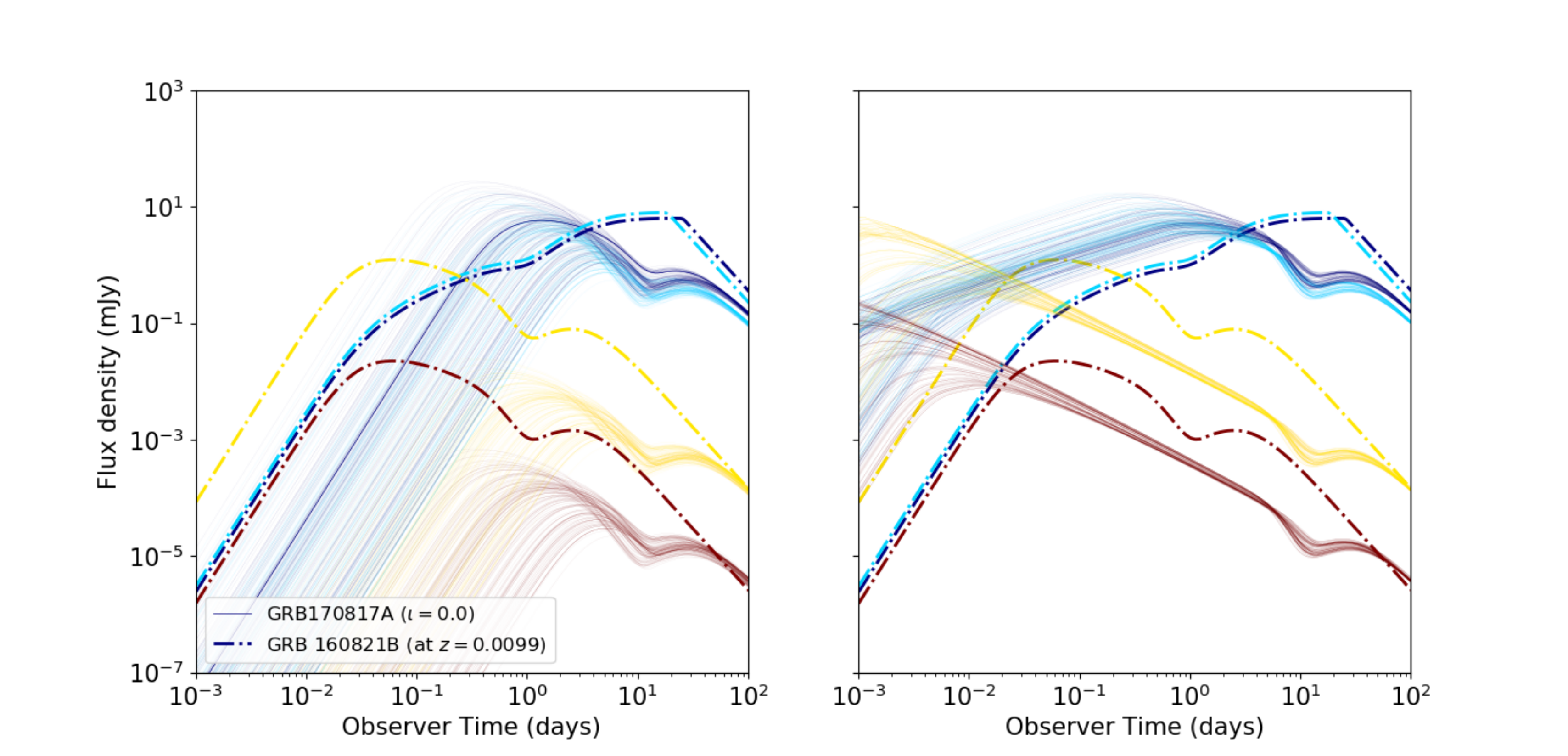}
    \caption{The on-axis view of the refreshed shock model fit to the GRB 170817A radio (3 GHz -- dark blue, 6 Ghz -- light blue), optical (F606W, $\sim5.1\times10^{14}$\,Hz -- yellow), and X-ray (1 keV -- red) frequencies. The solid lines show the light-curves using 400 randomly sampled posterior distribution parameter sets for GRB\,170817A but changing $\iota$ to $0.0$.  
    The afterglow from GRB\,160821B \citep{lamb2019b}, moved to the distance of GRB\,170817A, is shown for comparison (dot-dashed line).
    The left panel shows the afterglow lightcurves from a selection of the posterior sample for Model 1, and the right panel shows parameters drawn from the posterior sample for Model 2 -- the higher typical Lorentz factor in Model 2 is evidenced by the earlier and brighter peak for the initial shell. In both cases, the refreshed shock is consistently seen in the afterglow decline at $\gtrsim10$ days post peak.}
    \label{fig:GRB170817A-on-axis}
\end{figure*}

To check how the inferred afterglow of GRB\,170817A compares with an on-axis example, we plot the on-axis, $\iota=0.0$\,rad, predicted light-curve, using the same afterglow parameter sets, in Fig.\,\ref{fig:GRB170817A-on-axis}; where Model 1 is shown in the left panel and Model 2 in the right panel.
For the comparison, we also plot the equivalent afterglow for the GRB\,160821B parameters but shifted to the same luminosity distance and redshift as GRB\,170817A.
Compared to GRB\,160821B (dash-dotted lines) the on-axis afterglow to GRB\,170817A is typically fainter between $\sim0.1$\,days and $\sim 100$\,days.
The on-axis afterglow to Model 2 peaks earlier and at a higher flux than that for the GRB\,160821B parameters, this reflects the higher typical Lorentz factor found for the initial jet in Model 2.
The re-brightening due to the refreshed shock occurs at about the same time for both Models 1 \& 2, as expected from the similar $\Gamma_2$ distributions, but at a much later time when compared to the case in GRB\,160821B, this is to be expected from the lower secondary Lorentz factor values from the GRB\,170817A afterglow parameter fits, $\Gamma_{170817A} = 7.8(8.3)$ vs $\Gamma_{160821B}= 24$.
At typical short GRB cosmological distances, for Model 1 the predicted on-axis afterglow to GRB\,170817A would be faint although still consistent with the short GRB population, where only $\sim 34\%$ of {\it Swift}-identified short GRBs have detected optical afterglows\footnote{We note here that X-ray afterglow detection is more common, with $\sim74\%$ of the short GRB population.
However, X-rays are typically complicated by excess emission from internal and external plateaux, making the clear distinction between an external shock afterglow and some other afterglow emission complex. For completeness, \cite{fong2015} lists the fraction of short GRBs with a radio afterglow at $\sim 7\%$.} \citep{fong2015}.
Model 2 is also consistent with 
%that expected from 
the short GRB afterglow population, showing rapidly fading optical and X-ray emission.
For both models the radio afterglow for an on-axis observer peaks at $\sim 1$\,day.

\begin{table}
	\centering
	\caption{The \texttt{emcee} parameter distribution (central, 16th and 84th percentile limits) for the refreshed shock model fits to the radio, optical, and X-ray data shown in Fig. \ref{fig:GRB170817A}. Model 1 is for two discrete shells with uniform energy and Lorentz factor; Model 2 is where the secondary shell has a distribution of energy with velocity determined by the index $1-s$.	}
	\label{tab:GRB170817A}
	\begin{tabular}{lrr} % four columns, alignment for each
		\hline
		Parameter & Model 1 & Model 2 \\
		\hline
		\vspace{1mm}
		$\log$ [$E_1$ (erg)] & $51.51^{+1.07}_{-0.76}$ & $51.70^{+1.16}_{-0.81}$\\
		\vspace{1mm}
		$\log$ [$E_{\rm total}$ (erg)] & $52.62^{+1.16}_{-0.77}$ & $53.10^{+1.19}_{-0.80}$\\
		\vspace{1mm}
		$\Gamma_1$ & $19.54^{+44.04}_{-8.66}$ & $162.18^{+219.7}_{-122.1}$ \\
		\vspace{1mm}
		$\Gamma_2$ & $\geq7.79^{+1.14}_{-1.22}$ & $\geq8.27^{+2.13}_{-1.54}$\\
		\vspace{1mm}
		$\theta_j$ (rad) & $0.09^{+0.02}_{-0.01}$ & $0.11^{+0.03}_{-0.02}$ \\
		\vspace{1mm}
		$\iota$ (rad) & $0.28^{+0.06}_{-0.02}$ & $0.31^{+0.08}_{-0.05}$ \\
		\vspace{1mm}
		$\log \varepsilon_B$ & $-2.29^{+0.88}_{-1.82}$ & $-3.31^{+1.18}_{-1.18}$ \\
		\vspace{1mm}
		$\log \varepsilon_e$ & $-1.86^{+0.69}_{-1.17}$ & $-1.68^{+0.69}_{-1.19}$ \\
		\vspace{1mm}
		$\log\,[n$ (cm$^{-3}$)] &  $-2.94^{+1.29}_{-0.75}$ & $-2.54^{+0.96}_{-1.01}$ \\
		\vspace{1mm}
		$p$ & $2.16^{+0.01}_{-0.03}$ & $2.17^{+0.01}_{-0.01}$ \\
		\vspace{1mm}
		$s$ & N/A & $9.72^{+3.43}_{-2.19}$ \\
		\hline
	\end{tabular}
\end{table}

\section{Discussion}\label{sec:disc}
We have investigated the behaviour of a refreshed shock afterglow from a simple top-hat jet seen off-axis, and showed that it reproduces the observed temporal behaviour of the afterglow to GRB\,170817A.
In our refreshed shock model, as the initial outflow decelerates the second component will catch-up and energise the shock resulting in a re-brightening, or refreshing, of the afterglow emission.
We considered two models to describe the refreshed shock:
Model 1 has a single episode of energy injection, whereas Model 2 has a period of continuous energy injection until a maximum energy is achieved.
For an off-axis observer, the total refreshed shock energy will dominate the afterglow emission as the off-axis peak flux, at a fixed inclination, is $F_p\propto E$, the energy in the shock.
For an observer at a favourable inclination, the resulting afterglow will rise slowly to peak before declining as a typical, post-peak afterglow.
At higher inclinations, the light-curve will appear consistent with that from a top-hat jet with the parameters of the refreshed shock.

The energy in the jets, given by $E_j = E (1-\cos\theta_j)$ where $E$ is the on-axis isotropic equivalent energy, for GRB\,170817A when fit by the Model 1 refreshed shock is $E_{j,1} = 0.16^{+1.85}_{-0.13}\times10^{50}$\,erg in the initial outflow, and $E_{j,{\rm total}} = 0.19^{+2.83}_{-0.15}\times10^{51}$\,erg for the total final energy in the refreshed shock jets.
For Model 2 the jets energy is $E_{j,1} = 0.30^{+4.54}_{-0.25}\times10^{50}$\,erg in the initial outflow, and $E_{j,{\rm total}} = 0.37^{+5.52}_{-0.31}\times10^{51}$\,erg for the total final energy\footnote{This assumes that the system has identical bipolar jetted outflows.}. 
The values for the initial outflow are consistent with the median kinetic energy in the jets $\langle E_j\rangle\sim 0.8^{+2.5}_{-0.7}\times10^{50}$\,erg for the short GRB population \citep{fong2015}, and consistent with the energy found for the launched jet in GRB\,170817 by \cite{lazzati2020}.

For an on-axis observer, the refreshed shock re-brightening of the afterglow for a typical cosmological distance short GRB would be below the detection threshold in nearly all situations for Model 1, and rapidly fading to below detection thresholds for Model 2.
The GRB\,160821B is one of the closest confirmed on-axis short GRBs \citep{lamb2019b, troja2019} and shows evidence for a refreshed shock at late times.
Despite this GRB's proximity, at a redshift $z=0.16$, the broadband afterglow observations at the time of the refreshed shock episode were challenging. 
%using the energy and opening angle from the refreshed shock model for this GRB's afterglow, the energy in the jets is $E_{j,1} = 7.1\times10^{47}$\,erg and $E_{j,{\rm total}} = 1.1\times10^{49}$\,erg.
%The lower energies in this example are due to the small half-opening angle of the outflow.

\cite{lamb2019b} suggested continued or restarted central engine activity to explain the re-brightened afterglow episode in GRB\,160821B; this activity could be due to  late fallback material onto a disk, where the increasing mass of the disk triggers rapid accretion onto the central compact object and the conditions for a secondary outflow are created. %however, the 
The exact details of the central engine requirements are beyond the scope of the current work, although we note that in the case of GRB\,170817A, a magnetar can be ruled out \citep[see][for a detailed discussion]{ai2020}.
From the energy in the jet, and assuming that the X-ray emission lasting $\sim300$\,s post burst is due to energy dissipation within the lower-Lorentz factor second outflow, then the fallback mass can be estimated;
for GRB\,160821B, this fallback mass is $\sim2\times10^{-3}$\,M$_\odot$.
For GRB\,170817A, given the typically low-Lorentz factor for the second outflow episode, $\Gamma_{0,2}\gtrsim8$, then any energy dissipated within this secondary outflow is likely reabsorbed by the jet and will contribute to the total kinetic energy of the outflow \citep{lambkobayashi2016, matsumoto2020}.
In this scenario, no bright X-ray emission would have accompanied the secondary outflow, and given the inclination of the system to the line-of-sight, if this emission was beamed into a cone of $1/\Gamma_{0,2}$, then for cases where the Lorentz factor was sufficiently high that photons could escape the outflow, an X-ray plateau from this source would have been difficult to detect.
Alternatively, if the X-ray plateau in GRBs is a result of lateral jet structure and the viewing angle \citep{beniamini2020a, organesyan2020}, then the high opacity in a low-Lorentz factor outflow could fully suppress the emission.
Where the suppression due to the opacity and thermalisation is not fully efficient then an X-ray transient, or an X-ray burst may be produced.
However, if an X-ray transient is due to shock-breakout from a cocoon, then X-ray emission could accompany the low-$\Gamma$, refreshed shock scenario \citep{matsumoto2018}.

If the X-ray plateaux are an indication of continued central engine activity that injects energy into the jet, then refreshed shocks could be ubiquitous in the GRB population, however, only cases where the injected energy is much larger than the initial energy would result in observable afterglow variability.
A visual inspection of the X-ray light curves of {\em Neil Gehrels Swift Observatory} short GRBs suggests that of those with an X-ray afterglow, $>30\%$ have evidence for additional energy input at later times, most commonly in the form of a long lived plateau, persisting at bright levels for hundreds to thousands of seconds (with the majority at the shorter end), or in the form of X-ray flares. 
Such properties are also common in long GRBs, and are widely ascribed to additional energy injection. 

The origin of the prompt emission in GRB\,170817A is difficult to explain with an off-axis viewed typical short GRB \citep{kasliwal2017,lambkobayashi2018}, however, some component of the observed GRB could be due to emission from the mid-point of a structured jet \citep{ioka2019}.
The required core energy and Lorentz factor are at the upper-limits for the short GRB population, however, consistent with those from afterglow modelling assuming  a Gaussian structured jet \citep[e.g.][]{gill2018, lamb2019a, troja2019GW170817}.
For the refreshed shock models shown here; the top-hat jet structure, the initial on-axis isotropic equivalent kinetic energy $\sim10^{51}$\,erg, and the separation of $\iota/\theta_j\sim3$ for both models, suggests\footnote{If we assume a typical $\gamma$-ray efficiency $\eta_\gamma\sim0.15$, where $E_\gamma = E_1/(\eta_\gamma^{-1}-1)$, and following the method in \cite{ioka2019} and including the opacity for high energy photons as \cite{lambkobayashi2016}, we find an isotropic equivalent $\gamma$-ray energy for an observer at $\iota=0.28(0.31)$\,rad of $E_{\gamma, {\rm iso}}\sim 6.0\times10^{44}$\,erg and $\sim2.5\times10^{38}$\,erg for models 1 \& 2 respectively, with spectral peaks at $E_p\sim0.55(0.23)$\,keV, and well below the $E_{\gamma,{\rm iso}}\sim 5.35\times10^{46}$\,erg and $E_p\sim185\pm62$\,keV for GRB\,170817A.} 
that the prompt emission of GRB\,170817A is better explained as the shock breakout from a cocoon of material that is inflated by the successful, ultra-relativistic jet \citep{bromberg2018, duffell2018, gottlieb2018, pozanenko2018}.
Then GRB\,170817A is more a ``burst of $\gamma$-rays" than a `classical' GRB \citep{kasliwal2017}.

\begin{figure}
	\includegraphics[width=\columnwidth]{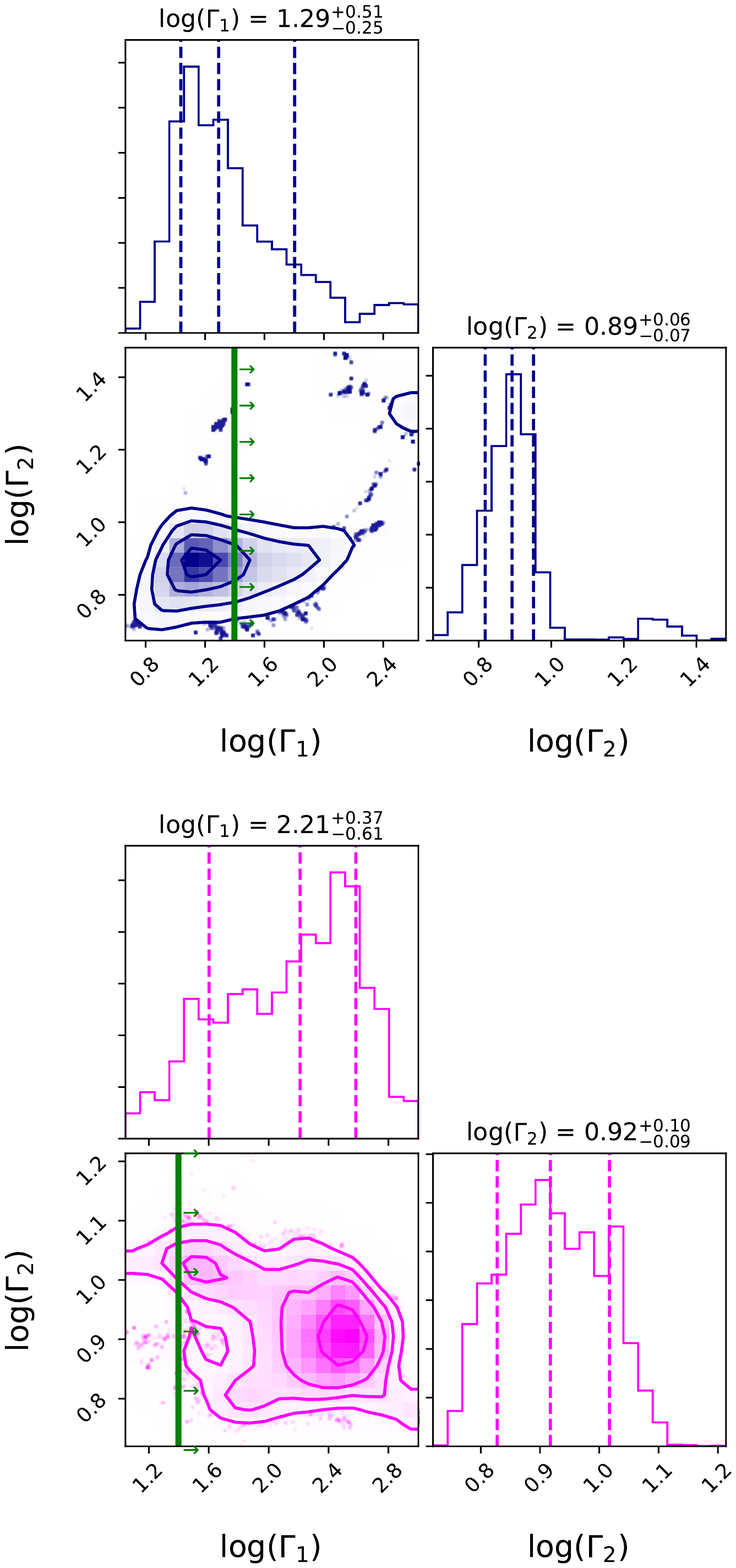}
    \caption{The 2D posterior distribution for the initial outflow Lorentz factor $\Gamma_{0,1}$ and the second outflow $\Gamma_{0,2}$. 
    The dashed vertical lines indicate the central, and the 16th and 84th percentiles.
    The green vertical line shows $\Gamma_{0,1}=25$, the $\sim$lower-limit for a successful GRB where the isotropic equivalent kinetic energy is $\sim$ $10^{51}$\,erg \citep{lambkobayashi2016}.
    Top panel (blue) shows posterior for Model 1.
    Bottom panel (pink) shows posterior for Model 2.}
    \label{fig:gamma-dist}
\end{figure}

An interesting result of the posterior distribution  from Model 1 for the Lorentz factor of the initial outflow is the preference of the model for `lower' values, $\Gamma_{0,1}=19.5^{+44.0}_{-8.7}$; see Fig.\ref{fig:gamma-dist}.
Given the energy of the initial outflow, $E_{1}=0.3^{+3.5}_{-0.3} \times10^{52}$\,erg, any dissipation within an outflow with $\Gamma_{0,1}\lesssim25$ is likely to be into an optically thick medium resulting in a failed-GRB \citep{dermer2000, huang2002}.
This implies that GRB\,170817A may be associated with a population of failed-GRBs that would result in on-axis orphan afterglows for all but the nearest events where the burst of gamma-rays from the shock-breakout may be detected.
Such a population could dominate the neutron star merger population, however, at cosmological distances the electromagnetic counterparts, even for an on-axis observer, would be difficult to detect \citep{lambkobayashi2016}.
However, if we restricted the prior range for the initial Lorentz factor to $\Gamma_{0,1}\gtrsim25$, then higher Lorentz factor solutions to the fit exist with some at $\Gamma\gtrsim100$, see the top (blue) panel in Fig.\,\ref{fig:gamma-dist}, although the clear preference of the model is for values $\Gamma_{0,1}\lesssim60$.
For Model 2, where the refreshing shock has a distribution of energy with velocity and results in a gradual injection of energy to the initial shell post collision, then the initial outflow has a Lorentz factor that is typically always above the condition for bright prompt emission for an on-axis observer; see the bottom (pink) panel in Fig.\,\ref{fig:gamma-dist}.

\begin{figure}
	\includegraphics[width=\columnwidth]{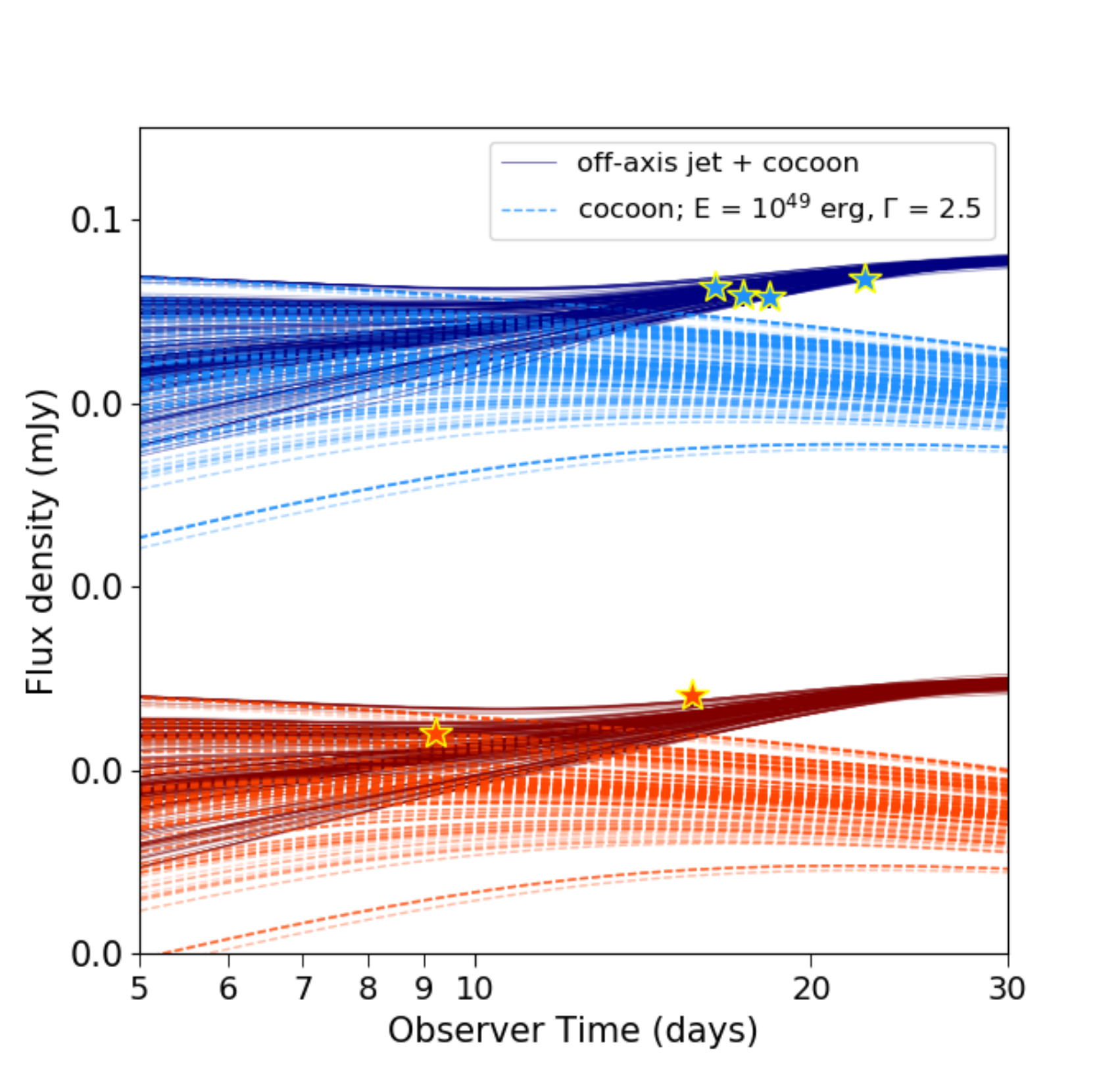}
	\caption{The light-curves from 400 randomly sampled posterior parameters sets from the refreshed shock Model 1 in Fig.\,\ref{fig:GRB170817A} are plotted with additional flux from an associated cocoon; solid lines. 
	The cocoon light-curve alone is shown as a dashed line.
	The cocoon has a fixed isotropic equivalent energy and Lorentz factor, $E=10^{49}$\,erg and $\Gamma=2.5$.
	In each case the cocoon surrounds the jet to an angle of $\theta_j + 20$\ang0.
	Red lines are the 1 keV flux and blue are the 3 GHz flux. 
	}
	\label{fig:cocoon}
\end{figure}

The rapid post-peak decline and the VLBI imaging of the afterglow to GRB\,170817A confirmed the ultra-relativistic jet origin for the emission \citep{lambmandelresmi2018, mooley2018, ghirlanda2019, hotokezaka2019}.
The temporal phenomenology of the afterglow's slow rise with $F\propto t^{\sim4/5}$ is usually attributed to the lateral structure of a core-dominated jet.
Although lateral structure is certainly present in a jetted outflow, the `steepness' of the jet edge structure is unknown \citep[although some simulations suggest very steep edges][etc.]{janka2006}, here we have shown that the afterglow of GRB\,170817A cannot conclusively be used to constrain the lateral jet structure and that the temporal features can be explained by a refreshed shock at late-times \citep[see also][for an alternative explanation using only a top-hat jet structure]{gill2019} -- similar to that seen in the afterglow to the short GRB\,160821B.
The earliest X-ray afterglow data at $\lesssim15$\,days are not well reproduced by the refreshed shock model. %, however, we note that this data is often `missed' by some structured jet models \citep[e.g.][]{lyman2018, margutti2018}.
However, we did not consider the emission from an associated mildly relativistic cocoon, where a cocoon surrounding the initial outflow could contribute to the emission at $t\lesssim10$\,days, \citep[e.g.][]{gottlieb2018a, salafia2020}.
In Fig.\,\ref{fig:cocoon} we show how the light-curves in Fig.\,\ref{fig:GRB170817A} are changed at $\lesssim10$\,days when we add a cocoon that surrounds each jet to an angle $\theta_j + 0.35$ rad.
The cocoon has a fixed isotropic equivalent energy $\sim10^{49}$\,erg, Lorentz factor $\Gamma=2.5$, and microphysical parameters $\varepsilon_B=0.001$ and $\varepsilon_e=0.1$; so the scatter between curves is a result of the different ambient densities.
From this, we can see that the missing flux at X-ray frequencies can be accounted for without any profile changes to the jet model.
A detailed investigation of the cocoon properties in GRB\,170817A is beyond the scope of this work.

We have focused our discussion on short GRB afterglows but the same arguments hold for long GRBs, where an orphan afterglow from a jetted refreshed shock outflow would be characterised by the total kinetic energy of the system $\sim10^{53}$\,erg, for long GRBs \citep{laskar2015}.
In cases where multiple injection episodes occur \citep[e.g.][]{granot2003}, the off-axis afterglow would, for an observer at a suitable inclination, appear to rise slowly to peak, or have a long plateau, with a maximum flux and timescale determined by the slowest and most energetic component.
The lengthened timescales for an orphan afterglow in such a case would increase the chance of detection for high cadence and sensitive transient survey telescopes.

\section{Conclusions}\label{sec:conc}

We have shown how the afterglow light-curve from a refreshed shock system for an observer at an angle to the jet axis $\iota>\theta_j$ would appear for a jet that has a simple top-hat structure.
We considered two models for the refreshed shock: (1) a discrete injection of energy at a given time and an assumed non-violent collision; and (2) an episode of continuous energy injection post-collision.
The key findings are:
\begin{itemize}
    \item The light-curves due to refreshed shock afterglows, when viewed by an off-axis observer, are dominated by the total (refreshed shock) energy; important when the kinetic energy of the secondary shell(s) is much greater than the first.
    \item A refreshed shock afterglow light-curve viewed at $\theta_j<\iota<\Gamma(t)^{-1}$ will show the same temporal features as those from a laterally structured jet resulting in orphan afterglows with longer, more gradual, rises to peak.
    \item The afterglow to GRB\,170817A can be described by a refreshed shock, top-hat jet viewed at $\sim3\theta_j$.
    
\end{itemize}

We find that the energy of the fast component in the afterglow model to GRB\,170817A data is consistent with the median kinetic energy from the cosmological sample of short GRB afterglows;
where the on-axis isotropic equivalent energy for Model 1(2) is $E_1 = 0.32(0.50)^{+3.48(6.74)}_{-0.26(0.23)}\times10^{52}$\,erg, and $\langle E\rangle = 0.18\times10^{52}$\,erg for the short GRB population \citep{fong2015}.
The inferred parameters from our MCMC for Model 1 indicate a bulk Lorentz factor with a preferred value $\Gamma_{0,1}= 19.5^{+44.0}_{-8.7}$, this indicates the possibility of a failed-GRB-type event \citep[e.g.][]{lambkobayashi2016}, alternatively, our Model 2 returns a Lorentz factor consistent with a bright on-axis GRB, with $\Gamma_{0,1}= 162.2^{+216.7}_{-122.1}$, where both models have an accretion driven secondary outflow \citep[e.g.][]{matsumoto2018, matsumoto2020}.
A low-Lorentz factor and/or an $\iota/\theta_j\sim3$ result in both scenarios being consistent with the hypothesis of a shock-breakout origin for the observed burst of gamma-rays, GRB\,170817A \citep{bromberg2018, pozanenko2018}.

\section*{Acknowledgements}

The authors thank the anonymous referee for helpful and constructive comments that have improved the paper.
G.P.L and N.R.T acknowledge support from STFC via grant ST/N000757/1.
A.J.L and N.R.T have received funding from the European Research Council (ERC) under the European Union's Horizon 2020 research and innovation programme (grant agreement No. 725246, TEDE, PI Levan).

%\vspace{5mm}
%\facilities{HST(STIS), Swift(XRT and UVOT), AAVSO, CTIO:1.3m,


%\software{astropy \citep{2013A&A...558A..33A},  
%          Cloudy \citep{2013RMxAA..49..137F}, 
%          SExtractor \citep{1996A&AS..117..393B}
%          }



\bibliography{ms}{}
\bibliographystyle{aasjournal}



\end{document}